
\def\gev{~{\rm GeV}}

\def\fbi{~{\rm fb}^{-1}}


\def\prdj#1{{\it Phys. Rev.} {\bf D{#1}}}
\def\npbj#1{{\it Nucl. Phys.} {\bf B{#1}}}

\def\plbj#1{{\it Phys. Lett.} {\bf B{#1}}}

\def\mt{m_t}
\def\rta{\rightarrow}
\def\tanb{\tan\beta}
\def\lplm{l^+l^-}
\input phyzzx
\Pubnum={$\caps UCD-92-26$\cr}
\date{November, 1992}

\titlepage
\vskip 0.75in
\def\nsd{N_{SD}}
\def\dg{\Delta g}
\baselineskip 0pt
\hsize=6.5in
\vsize=8.5in
\centerline{{\bf USING GLUON FUSION TO PROBE CP VIOLATION}}
\vskip .05in
\centerline{{\bf IN THE HIGGS SECTOR}\foot
{A shortened version will
appear in {\it Proceedings of the 1992 Division of Particles
and Fields Meeting}, Fermilab, Nov. 11-15 (1992), World Scientific.}
}
\vskip .075in
\centerline{John F. Gunion}
\vskip .075in
\centerline{\it Davis Institute for High Energy Physics}
\centerline{\it Department of Physics, U.C. Davis, Davis CA 95616}
\vskip .075in
\centerline{ABSTRACT}
\vskip .075in
\centerline{\Tenpoint\baselineskip=12pt
\vbox{\hsize=12.4cm
\noindent A technique for directly
probing CP violation in the Higgs sector of a multi-doublet
model using gluon-gluon collisions at the SSC is reviewed.}}
\vskip .15in
\noindent{\bf 1. Introduction}
\vskip .075in

Understanding the Higgs sector is one
of the fundamental missions of future high energy colliders such
as the SSC and LHC. In particular, it will be important to
know if CP violation is present in the Higgs sector. Generally,
either spontaneous or explicit CP violation can be present
if the Higgs sector consists of
more than the single doublet field of the Standard Model (SM).
\REF\hhg{J.F. Gunion, H.E. Haber, G. Kane and S. Dawson,
{\it The Higgs Hunter's Guide},
Frontiers in Physics Lecture Note Series \#80,
(Addison-Wesley Publishing Company, Redwood City, CA, 1990).}
(For a review of this and other issues summarized below, see
Ref.~[\hhg], and references therein.)
The presence of significant CP violation
would provide strong constraints on possible models.  For instance,
although CP violation is certainly possible in a general two-Higgs-doublet
model (2HDM), it does not arise (either explicitly or spontaneously)
in the highly restricted version of the 2HDM that
emerges in the Minimal Supersymmetric Model (MSSM).
Nor does CP violation occur spontaneously in the simplest extension of the MSSM
containing one additional singlet Higgs field and only trilinear terms in
the superpotential. In addition, in general supersymmetric models a large
CP-violating phase associated with the vacuum expectation value of a
Higgs field will give rise to large imaginary terms in the mass matrices
for the neutralino-gaugino fermions and the squarks.  The resulting imaginary
parts of the propagators for the mass eigenstates would be CP-violating and
result in contributions to neutron and electron electric dipole moments (EDM's)
that would exceed
current experimental limits.  Overall, observation of large CP violation
in the Higgs sector would be very difficult to reconcile with a
supersymmetric theoretical structure.  Thus, it will be of great
interest to determine the CP nature of any Higgs boson that is found.

Although there are a variety of experimental observables that are indirectly
sensitive to CP violation in the Higgs sector (such as EDM's, top quark
production and decay distributions, \etc), CP-violating contributions
typically first appear at one-loop, or are otherwise suppressed, and
will be very difficult to detect in a realistic
experimental environment.  In addition,
if CP violation in this class of observables is detected, it could easily
arise from sources other than the Higgs sector.  Consequently,
direct probes of CP violation in the Higgs sector are sorely needed.
Here we describe a {\it production rate}
asymmetry that directly probes the CP nature of any (observed) Higgs boson
that is produced via gluon-gluon fusion at the SSC/LHC.
(The proposed asymmetry is closely analogous to that developed previously for
collisions of polarized back scattered laser beams at a future
linear $\epem$ collider.
\Ref\bslaserbeasm{J.F. Gunion and B. Grzadkowski, preprint
UCD-92-18, \plbj{} (1992) in press.})
The primary experimental requirement is the ability to polarize the
colliding protons.
The difference between Higgs boson production rates for colliding beams
of different polarizations can be quite large in
a general 2HDM.  Asymmetries larger than 10\% are quite typical.
The reason that such large asymmetries can be achieved
is easily explained.  First, we note that
in the neutral Higgs sector a CP-violating phase for one of the neutral
field vacuum expectation values leads to mass matrix
mixing between the CP-even scalar fields and the CP-odd scalar fields.
Since the $gg$ coupling to CP-even scalars is not very different in
magnitude from that to CP-odd scalars (both arising at one loop, see below)
there is clear opportunity for large interference effects when the
Higgs mass eigenstate ($\phi$) contains substantial components of both types.
Of course, event rates in channels where $\phi$ can be detected
are not guaranteed to be large, and the observability of the predicted
maximal asymmetry must be carefully evaluated.

\smallskip
\noindent{\bf 2. Gluon Fusion}
\smallskip

The basic production mechanism in question is gluon fusion through a triangle
diagram. In gluon fusion, any colored particle which acquires its mass
via the Higgs mechanism can appear in the loop.
In particular, a heavy quark does {\it not} decouple when its mass is
much greater than the Higgs boson mass. Thus, in the SM
the top quark typically dominates. Additional arbitrarily heavy quarks
would each contribute coherently to the loop sum, implying
that models with additional
heavy families would yield far larger rates than obtained in the SM.
Of course, other types of colored particles appearing in extended models
can also contribute.
In the computations quoted here, we consider the model in which
the only extension of the SM occurs in the Higgs sector, but clearly the
rates and consequent detectability
of CP-violating asymmetries could be very much larger
in theories containing many heavy fermions.

At hadron colliders, the gluons within the hadrons provide a large
effective gluon-gluon collision luminosity.  The procedure for
computing the $gg\rta$Higgs cross section
in leading order is well-known.\refmark{\hhg}
Our computations will employ the leading order formalism, but it should
be noted that radiative corrections to this procedure
have been computed, and for a typical value of $\alpha_s$ result in
an enhancement factor of about 1.7.
\Ref\dawzer{ S. Dawson, \npbj{359}
(1991) 283; A. Djouadi, M. Spira, and P. Zerwas, \plbj{264} (1991) 440.}
In this sense, our results will be conservative.

Crucial to our discussion is the degree of polarization that
can be achieved for gluons at the SSC. The amount of gluon polarization
in a positively-polarized proton beam,
defined by the structure function difference $\dg(x)=g_+(x)-g_-(x)$,
is not currently known with any certainty.
(Here, the $\pm$ subscripts indicate gluons with $\pm$ helicity, and
 $g(x)=g_+(x)+g_-(x)$ is the unpolarized gluon distribution function.)
Many models used to describe the EMC data require a significant
amount of the proton's spin to be carried by the gluons. A
value for $\dg=\int\dg(x)\,dx$ of $\dg\sim 3$ is not atypical.  In the results
to be quoted, we shall employ the form:
\Ref\bergerqiu{E.L. Berger and J. Qiu, \prdj{40} (1989) 778.}
$\dg(x)=g(x)~~(x>x_c)$, $\dg(x)= {x\over x_c}g(x)~~(x<x_c)$,
where $x_c\sim 0.2$ yields a value of $\dg\sim 3-3.5$ over the Higgs
mass range (which determines the momentum transfer scale
at which $g(x)$ is evaluated) that we consider.
(Strictly speaking, $\dg(x)$ should be chosen to be of this
form at some given $Q_0$ and then evolved to obtain the form
at other $Q$ values --- we ignore this subtlety here.)  Although this
form maximizes $\dg(x)$ at large $x$, the $x$ values most important for
Higgs production are substantially below 0.2.

\smallskip
\noindent{\bf 3. Asymmetry and Results}
\smallskip

The asymmetry we compute is simply
$A\equiv[\sigma_+-\sigma_-]/[\sigma_++\sigma_-]$,
where $\sigma_{\pm}$ is the cross section for Higgs production in collisions
of an unpolarized proton with a proton of helicity $\pm$, respectively.
$\sigma_+-\sigma_-$ is proportional to the integral over $x_1$ and $x_2$
(with $x_1x_2=m_\phi^2/s$) of
$g(x_1)\dg(x_2)\left[|{\cal M}_{++}|^2-|{\cal M}_{--}|^2\right]$,
while $\sigma_++\sigma_-$ is determined by the integral of
$g(x_1)g(x_2)\left[|{\cal M}_{++}|^2+|{\cal M}_{--}|^2\right]$.
(We have assumed that it is proton 2 that is polarized.) Now,
$|{\cal M}_{++}|^2-|{\cal M}_{--}|^2$ vanishes for
a CP eigenstate, but can be quite large in a general 2HDM; the difference
is proportional to ${\rm Im\,}(eo^*)$, where $e$ ($o$) represents the
$gg$ coupling to the CP-even (-odd) component of $\phi$.

To obtain a numerical indication of the observability of $A$, we have
proceeded as follows. First, we maximize
$(\sigma_+-\sigma_-)/\sqrt{\sigma_++\sigma_-}$ (which determines
the statistical significance of our observation) by choosing an optimum
value for $x_F^{cut}$ such that only $x_2-x_1>x_F^{cut}$ is included
in our integrals; such a cut pushes the integrals
into the region where $\dg(x_2)/g(x_2)$ is less suppressed ---
$x_F^{cut}$ near 0.05 typically yields optimal results.
Second, we assume that $\phi$ can only be
detected in the $\phi\rta ZZ\rta \lplm X$ modes. Third, we have
searched (at fixed $\tanb=v_2/v_1$) for the
parameters of the most general CP-violating 2HDM that yield
the largest achievable statistical significance, $\nsd^{max}$, for measuring
$A$ in the $ZZ\rta\lplm X$ mode ---
the need for keeping a large rate in the $ZZ$
mode competes with the requirement of keeping ${\rm Im\,}(eo^*)$ large.

\FIG\ggcphiggsnsd{}
\midinsert
\vbox{\phantom{0}\vskip 4.5in
\phantom{0}
\vskip .5in
\hskip -20pt
\special{ insert scr:ggcphiggs_nsd.ps}
\vskip -1.4in }
\centerline{\vbox{\hsize=12.4cm
\Tenpoint
\baselineskip=12pt
\noindent
Figure~\ggcphiggsnsd:
Maximal statistical significance achieved for asymmetry signal as
a function of $m_\phi$ at the SSC with $L=10\fbi$. The branching ratio
for $\phi\rta ZZ\rta\lplm X$ is included.
}}
\endinsert

The results for $\nsd^{max}$ at the SSC with integrated
luminosity of $10\fbi$ appear in Fig. \ggcphiggsnsd, for $\tanb=2$ and $10$,
and $\mt=150\gev$.
Detection of this asymmetry is clearly not out
of the question.  Indeed, the Higgs sector parameters required
to achieve the illustrated $\nsd^{max}$ are not at all fine tuned.
Large ranges of parameter space yield values very nearly as big.
It should be noted that for the parameter choices which yield these maximal
results, the production rate for $\phi$ at $\tanb=2$ is quite similar
to that for production of a SM Higgs boson of the same mass, once
$m_\phi>2\mt$, and that
the branching ratio for $\phi\rta ZZ,WW$ decays is close to the SM value.
Thus, without this asymmetry measurement, distinguishing between
a $\phi$ which is a CP mixture and the SM Higgs boson could be difficult.
At $\tanb=10$, the $\phi$ production rate becomes increasingly suppressed
relative to the SM value at larger $m_\phi$,
and would alone indicate a non-SM scenario.

Our ability to detect $A$ may be either better or worse than that illustrated
in Fig. \ggcphiggsnsd. If only partial polarization for
the proton beam can be achieved $\nsd^{max}$
would worsen. Limited acceptance for the final states of interest
would also cause $\nsd^{max}$ to decrease by a factor of $\sqrt{\epsilon}$,
where $\epsilon$ is the acceptance efficiency.
If additional decay modes could be employed
(\eg\ $WW$ decay channels with one $W$ decaying leptonically)
or if enhanced luminosity can be used,
$\nsd^{max}$ could be improved.
Of course, if only the $ZZ\rta \lplm\lplm$ channel can be employed then the
statistical significance is decreased (by a factor of roughly 5.5);
a higher luminosity would definitely be required to detect the asymmetry.
However, we are optimistic that once a Higgs boson is found and its mass
known, techniques for employing more than just the gold-plated $4l$
mode will be found.

\smallskip
\noindent{\bf 4. Conclusion}
\smallskip

The ability to polarize one of the proton beams at the SSC will provide
a unique opportunity for determining the CP nature of any observed neutral
Higgs
boson.  Indeed, if the Higgs boson has both significant CP-even and CP-odd
components, then a large asymmetry between production rates for
positively versus negatively polarized protons will arise.
If measurable CP violation is found in the Higgs sector
many otherwise very attractive models will be eliminated, including
the Standard Model and most supersymmetric models.
This should provide a rather strong
motivation for expending the relatively modest monetary amounts
needed to achieve polarized SSC beams.

\smallskip\noindent{\bf 5. Acknowledgements}
\smallskip
The results described here were obtained
in collaboration with T.-C. Yuan and B. Grzadkowski.
This work has been supported in part by the Department of Energy.
\smallskip
\refout
\end